\documentclass[10pt]{iopart}
\usepackage{graphicx}
\usepackage{morefloats}
\usepackage{setstack}
\usepackage{color}
\usepackage{epsf,epsfig,graphics}
\def\etal{{\it et al~}}

\begin{document}

\title[W$^{37+}$ - W$^{28+}$ DR rate coefficients]{Dielectronic recombination of the open 
$4d$-shell of Tungsten: W$^{37+}$ -- W$^{28+}$}

\author{S. P. Preval}\author{N. R. Badnell}\author{M. G. O'Mullane}

\address{Department of Physics, University of Strathclyde, Glasgow G4 0NG, United Kingdom}
\ead{simon.preval@strath.ac.uk}
\vspace{10pt}
\begin{indented}
\item[]November 2016
\end{indented}

\begin{abstract}
Tungsten is an important element for magnetically confined fusion plasmas but has the potential to cool, 
or even quench the plasma due to it being an efficient radiator. Total and level-resolved dielectronic 
recombination (DR) rate coefficients, for all ionization stages, are essential to model tungsten. We 
describe a set calculations performed using the distorted wave code {\sc autostructure} for the tungsten 
ions W$^{37+}$ to W$^{28+}$. We demonstrate the importance of relativistic configuration mixing in such calculations. 
In particular, we show that the partial DR rate coefficients calculated in level and configuration resolution 
can differ by as little as 5\%, and up to as much as 75\%. Using the new data, we calculate a revised 
steady-state ionization fraction for tungsten. We find that, relative to the ionization fraction calculated 
using the recombination rate coefficients of Putterich~\etal (Plasma Phys. Control. Fusion, 50, 085016), 
the peak temperatures of W$^{37+}$ to W$^{28+}$ ionization states are shifted to lower temperatures spanning 
0.9-1.6keV. This temperature range is important for understanding the performance of large tokamaks, 
such as ITER, because the temperatures in the pedestal, edge, scrape-off-layer and divertor region fall 
in this range.

\end{abstract}

%
%
%
%
\ioptwocol

\section{Introduction}
The experimental thermonuclear tokamak reactor ITER is currently being built in Cadarache, 
France, with its first plasma projected to be in 2025. Designed to output 10 times as much 
energy as it consumes, it is posited to be the penultimate step in realising a commercial fusion 
reactor. The plasma facing component of the the divertor will be coated with tungsten. This 
metal has been chosen due to its resistance to tritium absorption, its ability to withstand 
large power loads, and its high melting point. However tungsten will be sputtered into the 
core, confined, plasma. Tungsten as an impurity in the confined plasma is a problem because 
it is a very efficient radiator, which will lead to cooling of the plasma, and potentially quenching 
it. A key method in understanding the impact of tungsten impurities on the tokamak plasma is 
detailed collisional-radiative modelling \cite{summers2006a}. However, such modelling requires 
the provision of partial, final-state resolved dielectronic/radiative recombination (DR/RR) rate 
coefficients for the ion being modelled.

High temperatures in the core are required to enable the fusion reaction and maintaining this 
core temperature necessitates a high temperature at the plasma edge. The pedestal region is 
characterized by a steep pressure ($p=n_{e}T_{e}$) gradient at the edge which forms a transport 
barrier; maximizing its height, and inter alia the edge temperature, is the preferred way to achieve 
high performance [2]. Unfortunately the plasma conditions of this high confinement (H-mode) 
regime are susceptible to MHD instabilities which collapse the pedestal, thus reducing the 
temperature and allowing impurities into the core region, resulting in reduced performance or 
destruction of the conditions needed to sustain the fusion reaction. Radiation from high efficiency 
radiators such as tungsten will affect the balance between the competing effects. For ITER the 
pedestal temperature is 1­4keV depending on model and extrapolations from JET ITER-like wall 
experiments \cite{kotschenreuther2017a,snyder2011a}. The ionization stages of tungsten reported 
here fall in this region so it is important to be able to model their distribution, particularly in regions 
like the pedestal, where turbulent transport is strongly suppressed.

Several isonuclear datasets for tungsten have been calculated using a variety of methods. The first 
was generated by Post~\etal \cite{post1977a,post1995a} using an average-ion method with the 
ADPAK codes. Later, P\"{u}tterich~\etal \cite{putterich2008a} used these recombination rate coefficients in 
modelling the tokamak plasmas from ASDEX upgrade, but scaled several ionization stages 
(W$^{20+}$--W$^{55+}$) by empirically determined constants to match observed spectral emission. 
Next, Foster \cite{foster2008a} used a combination of the Burgess General Formula \cite{burgess1965a} and the Burgess-Bethe
General Program \cite{badnell2003a} to calculate DR rate coefficients for the sequence. The RR rate coefficients were calculated
by scaling hydrogenic results. Lastly, there exists a complete set of recombination rate coefficients on the
International Atomic Energy Agency website \footnote{https://www-amdis.iaea.org/FLYCHK/} calculated 
using the FLYCHK code \cite{chung2005a}, which also uses an average-ion method. Despite the numerous
calculations available and the wide range of methods used, poor agreement between the calculated recombination
rate coefficients is observed between all three datasets. Examples of this poor agreement can be 
seen in Kwon~\etal \cite{kwon2017a}, where multiple comparisons are made between these datasets. 
Clearly, further theoretical investigation is required to resolve this disagreement.

The demand for partial final-state resolved DR rate coefficients for tungsten has seen 
a surge in the number of calculations being performed. Typically, these calculations covered ions 
where the outer valence electron shell is full or nearly full, which are useful for plasma 
diagnostics. Safronova~\etal has calculated DR rate coefficients for a wide range of tungsten ions 
using the Cowan \cite{cowanbook1981} and {\sc hullac} \cite{barshalom2001a} codes, covering 
W$^{5+}$, W$^{6+}$, W$^{28+}$, W$^{38+}$, W$^{45+}$, W$^{46+}$, W$^{63+}$, and W$^{64+}$ 
\cite{usafronova2012a,usafronova2012b,usafronova2011a,usafronova2016a,
usafronova2015a,usafronova2012c,usafronova2009a,usafronova2009b}. Behar~\etal also used the Cowan
and {\sc hullac} codes to calculate DR rate coefficients for W$^{45+}$, W$^{46+}$, W$^{56+}$, and W$^{64+}$ 
\cite{behar1997a,behar1999b,behar1999a}. The Flexible Atomic Code ({\sc fac}) can also calculate DR rate coefficients,
and was used by Li~\etal to calculate data for W$^{29+}$, W$^{39+}$, W$^{27+}$, W$^{28+}$, and W$^{64+}$ 
\cite{li2012a,li2014a,li2016a}. In addition, Meng~\etal and Wu~\etal have also used {\sc fac} to calculate DR for
W$^{47+}$, and W$^{46+}$-W$^{46+}$ respectively. Lastly, Kwon~\etal used {\sc fac} to calculate DR rate coefficients
for W$^{44+}$-W$^{46+}$ \cite{kwon2016a,kwon2016b}.

Several works are now exploring ions of tungsten where the valence shell is partially or half full. 
The most complicated example to date is the work of Spruck~\etal and Badnell~\etal 
\cite{spruck2014a,badnell2016a,badnell2012a}, where the authors compared experimental storage ring 
measurements of DR for W$^{18+}$, W$^{19+}$, and W$^{20+}$ (ground configurations $4f^{10}$, 
$4f^{9}$, and $4f^{8}$ respectively) with calculations from the atomic collision package 
{\sc autostructure} \cite{badnell1986a,badnell1997a,badnell2011a}. Other approaches to dealing with 
such complex ions include the use of statistical methods such as those employed by Dzuba~\etal 
\cite{dzuba2012a,dzuba2013a}, Berengut~\etal \cite{berengut2015a}, and Harabati~\etal \cite{harabati2017a}.
An excellent review of the methods used to study $4f$-shell ions, both experimentally and theoretically, 
has been compiled by Krantz~\etal \cite{krantz2017a}.

The most recent attempt to cover the entire isonuclear sequence is known informally as 
\textit{The Tungsten Project}. Using {\sc autostructure}, Preval~\etal has calculated DR and 
RR rate coefficients for W$^{73+}$--W$^{38+}$ \cite{preval2016a,preval2017b}. 
Because of the increased activity in calculating data for tungsten, multiple comparisons have been
performed with data from \textit{The Tungsten Project}, confirming the reliability of the calculation
methods used. All data from these publications have been published on the OPEN-ADAS 
website\footnote{{h}ttp://open.adas.ac.uk} in the standard adf09 (DR) and adf48 (RR) formats. The
definitions of these formats can also be found on the OPEN-ADAS website.

While not as complicated as the open $4f$ shell, calculating DR rate coefficients for open $4d$ 
shell ions still represents a significant jump in complexity to ions considered previously. The 
rich level structure of these ions also presents interesting physics in its own right. It is 
for this reason that this paper is dedicated to the $4d^{q}$ ($q=1-10$) isonuclear sequence of 
tungsten, covering charge states W$^{37+}$--W$^{28+}$. As in Preval~\etal \cite{preval2016a,preval2017b} 
we use some technical notation when discussing the various ionization stages. Beyond Zn-like, referring to 
ions by their isoelectronic chemical symbol requires a good memory of the periodic table. It is for this reason 
that we refer to the various ionization stages by the number of valence electrons instead. For example, H-like 
($Z=1$) becomes 01-like, and Pd-like ($Z=46$) with 46 valence electrons becomes 46-like.

We structure this paper as follows. We first describe the underlying theory for our DR 
calculations. Next, we present and discuss our results, and compare our DR rate coefficients to 
those calculated in other works. Finally, we assess the quantitative effect of using our data in 
calculating the steady state ionization fraction for tungsten, and demonstrate how the $4d^{q}$ 
ions can be potentially used as a temperature diagnostic. We conclude with a few remarks.

\section{Theory} \label{theorysec}
The theory underpinning the calculation of DR rate coefficients has been discussed at length in 
multiple works, however, we summarise the basic principles here. All DR rate coefficients described 
in this paper were calculated using the distorted wave code {\sc autostructure}. It is able to 
calculate energy levels, radiative/autoionization rates, collision strengths, and many other quantities. The code
is able to calculate data in configuration average (CA), term (LS), or level (IC) resolution using
relativistic $\kappa$-averaged wavefunctions. {\sc autostructure} has been benchmarked extensively 
against experimental DR measurements for various ions such as Fe$^{19+}$ \cite{savin2005a}, and 
F$^{5+}$ \cite{ali2013a}.

For an ion $X_{\nu}^{(+z+1)}$ with residual charge $z$ in an initial state $\nu$ recombining into an ion 
$X_{f}^{+z}$ with final state $f$, the partial DR rate coefficient $^{DR}\alpha_{f\nu}^{z+1}$ at a 
particular electron temperature, $T_e$, can be written as
\begin{eqnarray}
^{DR}\alpha_{f\nu}^{z+1}(T_e)&=&\left(\frac{4{\pi}a_{0}^{2}I_{H}}{k_{B}T_{e}}\right)^{\frac{3}{2}}
                                \sum_{j}\frac{\omega_{j}}{2\omega_{\nu}}\exp{\left[-\frac{E}{k_{B}T_{e}}\right]} \nonumber \\
                                &\times&\frac{\sum_{l}{A_{j\rightarrow{\nu},E\,l}^a}{A_{j\rightarrow{f}}^{r}}}
                                {\sum_{h}{A_{j\rightarrow{h}}^{r}} + \sum_{m,l}{A_{j\rightarrow{m},E\,l}^{a}}},
\label{DReq}
\end{eqnarray}
where $\omega_{\nu}$ and $\omega_{j}$ are the statistical weights of the $N$- and $(N+1)$-electron ions 
respectively, the $A^{r}$ and $A^{a}$ are the radiative and Auger rates respectively, and $E$ is 
the total energy of the continuum electron minus its rest energy, which is fixed by the position 
of the resonances. The sum over $l$ is performed over all orbital angular momentum quantum numbers 
included, and the sum over $j$ is over all autoionizing levels. The sums over $h$ and $m$ give the 
total radiative and Auger widths respectively. Lastly, $I_{H}$ is the ionization energy of the 
hydrogen atom, $k_B$ is the Boltzmann constant, and $(4{\pi}a_{0}^{2})^{3/2}=6.6011\times{10}^{-24}$cm$^{3}$.

The RR rate coefficients for the $4d$ ions are now completely overwhelmed by the DR rate cofficients.
Therefore, they are calculated for \textit{The Tungsten Project} for completeness only. RR contributes at most
$\sim{4}$\% for 37-like at peak abundance temperature ($2.7\times{10}^{7}$K), decreasing to $\sim{1}$\%
for 46-like at peak abundance temperature ($2.7\times{10}^{7}$K). The RR rate coefficient 
$^{RR}\alpha_{f\nu}^{z+1}(T_e)$ can be written in terms of its inverse, photoionization $^{PI}\sigma_{\nu f}^{z}(E)$, 
using detailed balance. Therefore, the partial RR rate coefficient can be written as:

\begin{eqnarray}
^{RR}\alpha_{f\nu}^{z+1}(T_e) &=& \frac{c\,\alpha^3}{\sqrt{\pi}}
                                  \frac{\omega_{f}}{2\omega_{\nu}}\left(I_H k_B T_e\right)^{-3/2} \nonumber \\
                                  &\times&\int^\infty_0 E^2_{\nu f}\, {^{PI}\sigma_{\nu f}^{z}(E)}
                                  \exp{\left[-\frac{E}{k_{B}T_{e}}\right]}dE\,,
\label{RReq}
\end{eqnarray}
where $E_{\nu f}$ is the corresponding photon energy, and $c\,\alpha^3/\sqrt{\pi}=6572.67$cm\,s$^{-1}$.

At high energies and temperatures, relativistic effects must be factored into the Maxwell-Boltzmann distribution.
This results in a correction factor to the distribution giving the Maxwell-J\"{u}ttner distribution \cite{synge57a},
which can be written as:
\begin{equation}
F_{\mathrm{r}}(\theta) = \sqrt{\frac{\pi\theta}{2}} 
\frac{1}{K_{2}(1/\theta){\rm e}^{1/\theta}},
\end{equation}
where $\theta=\alpha^2 k_{B}T/2I_H$, $\alpha$ is the fine-structure constant and $K_{2}$ is the modified Bessel function of 
the second kind. As we move along the isonuclear sequence the J\"{u}ttner correction becomes less
relevant. For the $4d$ ions considered in the present work, the J\"{u}ttner correction causes a change of
$<1$\% at peak abundance for 37-like, and even less for 46-like. However, the high-temperature
DR decreases by as much as 50\% for 37-like at $2.74\times{10}^{9}$K, and 35\% for 46-like at $1.57\times{10}^{9}$K.

\section{Calculations}

\subsection{DR}
We split our DR rate coefficient calculations into core excitations, which are labelled in terms of 
the initial and final principal quantum numbers $n$ and $n'$ of the promoted target electron. The 
largest contributions to the total DR rate coefficient by far come from the outer shell 
$\Delta{n}=0,1$ core excitations 4--4 and 4--5. In the case of the inner shell $\Delta{n}=1$ core 
excitation (3--4), the filling of the $n=4$ shell restricts the number of promotions that can occur 
from $n=3$, decreasing its importance. The converse is true for the outer shell $\Delta{n}=2$ core 
excitation (4--6). 

The 3--4 core excitation was calculated in CA due to its small contribution to the recombination
rate coefficient total. The $N$-electron configuration set included all
possible double excitations of $3\ell$ and $4\ell$ electrons from the ground configuration of the ion
considered. The $(N+1)$-electron configurations consisted of all possible triple excitations of $3\ell$
and $4\ell$ electrons from the ground configuration of the $(N+1)$-electron ion.

The 4--6 core excitation was also calculated in CA due to its small contribution to the recombination
rate coefficient total. The $N$-electron configuration set was constructed in a similar manner to
the 3--4 core excitation, and included all possible double excitations of the $4\ell$ electrons from the ground configuration 
to $5\ell$ and $6\ell$. The $(N+1)$-electron configurations included all possible triple excitations of 
the $4\ell$ electrons from the $(N+1)$ ground configuration to $6\ell$.  

The 4--4 core excitation was calculated in IC due to its large contribution to the recombination rate
coefficient total. The $N$-electron configurations consisted of single promotions from $4p$
and $4d$ to $4\ell$ from the ground configuration of the ion being considered. Promotions from $4s$ were 
omitted due to their small contribution to the DR rate coefficient total. Mixing configurations 
were included by way of the ``one-up one-down'' rule as described in Cowan's book 
\cite{cowanbook1981}. For example, for a particular configuration $4p^{\mu} 4d^{\nu} 4f^{\rho}$ the 
corresponding mixing configurations would be:
\begin{itemize}
\item $4p^{\mu-1} 4d^{\nu+2} 4f^{\rho-1}$
\item $4p^{\mu+1} 4d^{\nu-2} 4f^{\rho+1}$.
\end{itemize}
The $(N+1)$-electron configurations consisted of the $N$-electron configurations with an additional
target electron added. We demonstrate the resulting configuration set for 40-like in Table \ref{table:configs40like}.
Given such a set of configurations, AUTOSTRUCTURE calculates all possible autoionization and
electric dipole radiative rates between them.

The 4--5 core excitation was also calculated in IC. The $N$-electron configuration set included 
promotions from $4p$ and $4d$ to $4\ell$ and $5\ell$. As with 4--4, promotions from $4s$ were 
omitted due to their negligible contribution to the total. Mixing configurations were included as 
described above. The 4--5 core excitation was the most computationally demanding calculation  
due to the large number of levels. To make the calculations tractable, promotions from $4p$ were 
omitted from 40- to 43-like, while ``one-up one-down'' mixing configurations were omitted from 39- 
and 45-like. As with 4--4, the $(N+1)$-electron configuration set is just the $N$-electron 
configuration set plus an additional target electron.

For each core excitation, DR from Rydberg $n\ell$ electrons were calculated sequentially up to 
$n=25$, and then on a quasi-logarithmic grid of $n$ values up to $n=999$. Interpolation was used 
to obtain the intermediate $n$ values. We included sufficiently many $\ell$ values so as to 
numerically converge the total DR rate coefficient to $<1$\% over the entire ADAS temperature range, 
spanning $z^{2}(10-10^7)$K. The number of $\ell$ values required to achieve numerical
convergence varies according to the ionization state, and the core excitation considered. The maximum
required was for 41-like 4--4 with $\ell=15$.

\subsection{RR}
RR is most important in the case of highly charged ions. For example, in the case of 02-like tungsten, 
RR contributes 100\% of the recombination rate total at 
peak abundance temperature (see \cite{preval2016a}). In the present work, we have calculated total
RR rate coefficients including contributions for dipole radiation only. This is because 
higher-order multipolar radiation contributions are manifest at high energies, and hence only higher 
ionization states of tungsten. Omission of these contributions leads to a change in the total
recombination rate coefficient of $\ll{1}$\% for 37-like at peak abundance temperature. Our $N$-electron 
configurations consisted of $4d^q$ and $4d^{q-1}4f$ ($q=1-10$). The $(N+1)$-electron configurations were 
just the $N$-electron configurations plus an additional Rydberg electron. As with DR, RR contributions 
from Rydberg $nl$ electrons were calculated sequentially up to $n=25$, and then 
quasi-logarithmically up to $n=999$, with interpolation being used to obtain the intermediate 
values. The Rydberg RR was calculated for sufficiently many $\ell$ values so as to numerically 
converge the RR rate coefficient to $<1$\% over the ADAS temperature range. Given its small
importance to the calculations, we will not discuss RR any further in this paper.

\section{Results}
We have calculated level- and configuration-resolved DR rate coefficients for 37- to 46-like 
tungsten. We split our discussion with respect to the core-excitations that have been calculated.

\subsection{3--4}
\begin{figure}
\begin{centering}
\includegraphics[width=85mm]{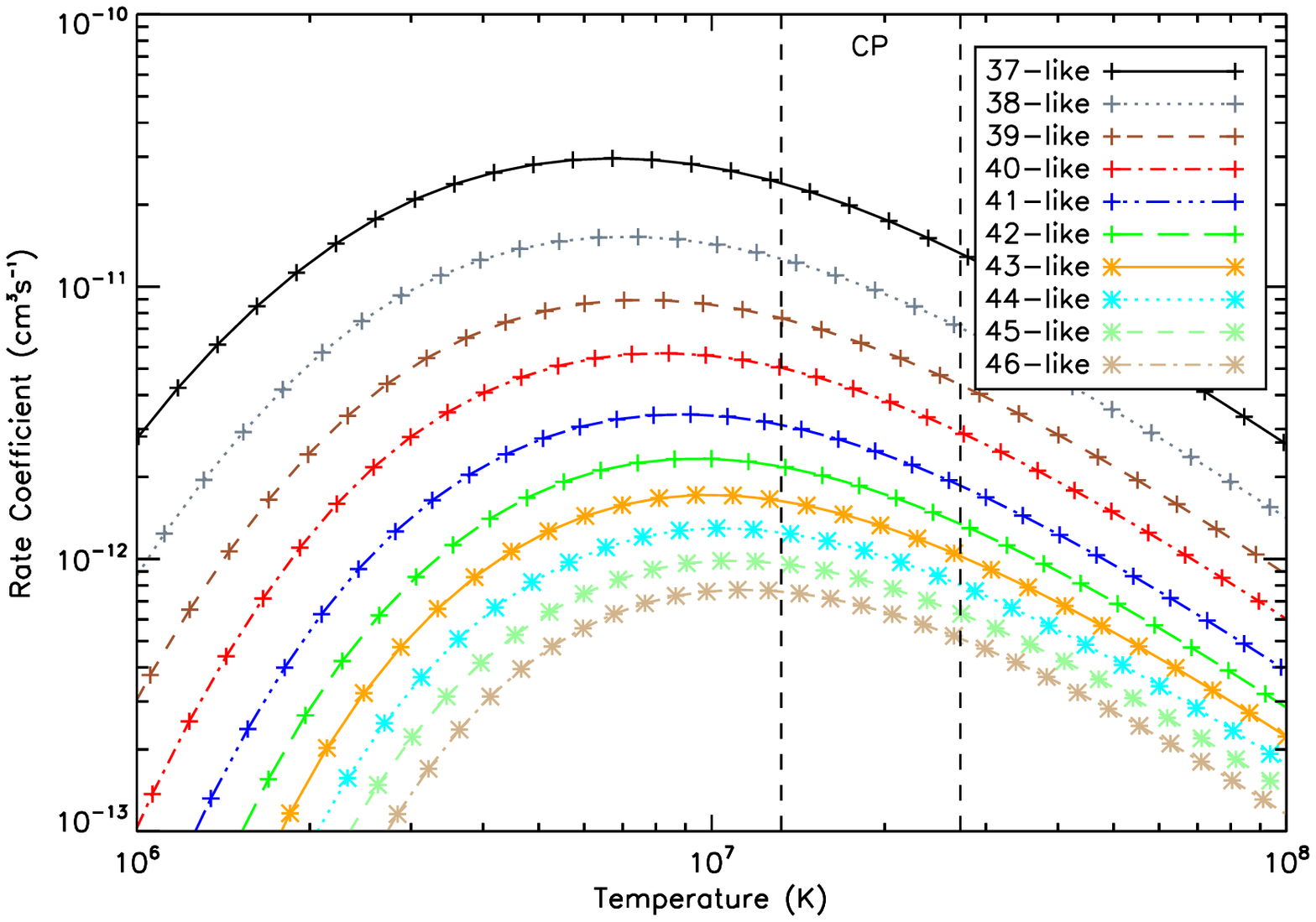}
\caption{Total DR rate coefficients for the 3--4 core excitation for 37- to 46-like, calculated
in CA.}
\label{fig:hcar34}
\end{centering}
\end{figure}

The 3--4 core excitation has been calculated in configuration average, and has been included for 
completeness. In Figure \ref{fig:hcar34} we have plotted the total 3--4 DR rate coefficients for 
37- to 46-like. We have also plotted two vertical lines on this figure labelled ``CP'' (collisionally 
ionized plasma). This range is defined as being the range of temperatures going from the peak 
abundance temperature of 37-like, down to the peak abundance temperature of 46-like. Note that 
the actual range of temperatures of interest will be wider than this. It can be easily seen that 
the total DR-rate coefficient decreases steadily as more $4d$ electrons are added. As the $4d$ shell 
is filled, fewer $3p\rightarrow{4d}$ electric dipole transitions can take place, decreasing the overall 
total. For 37-like 3--4 contributes $\sim{15}$\% to the total for temperatures 
$>{3\times{10}^8}$K, but only $\sim{12}$\% at peak abundance temperature 
($2.7\times{10}^{7}$K). As we move along the isonuclear sequence, this contribution rapidly drops 
to $<{1}$\% of the total recombination rate coefficient by 46-like at peak abundance.

\subsection{4--4} \label{corex44}
\begin{figure}
\begin{centering}
\includegraphics[width=85mm]{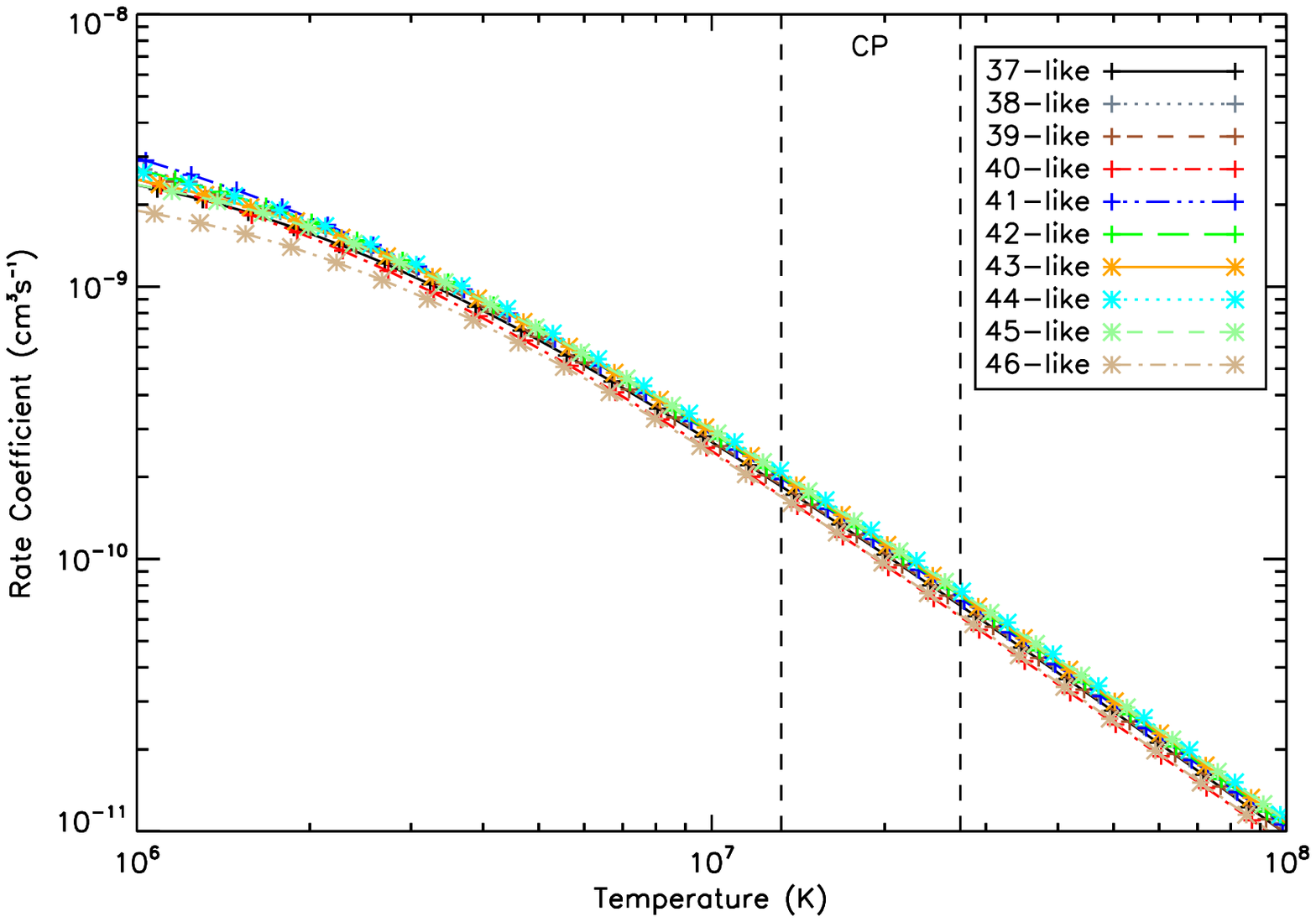}
\caption{Total DR rate coefficients for the 4--4 core excitation for 37- to 46-like, calculated
in IC.}
\label{fig:hicr44}
\end{centering}
\end{figure}

\begin{figure}
\begin{centering}
\includegraphics[width=85mm]{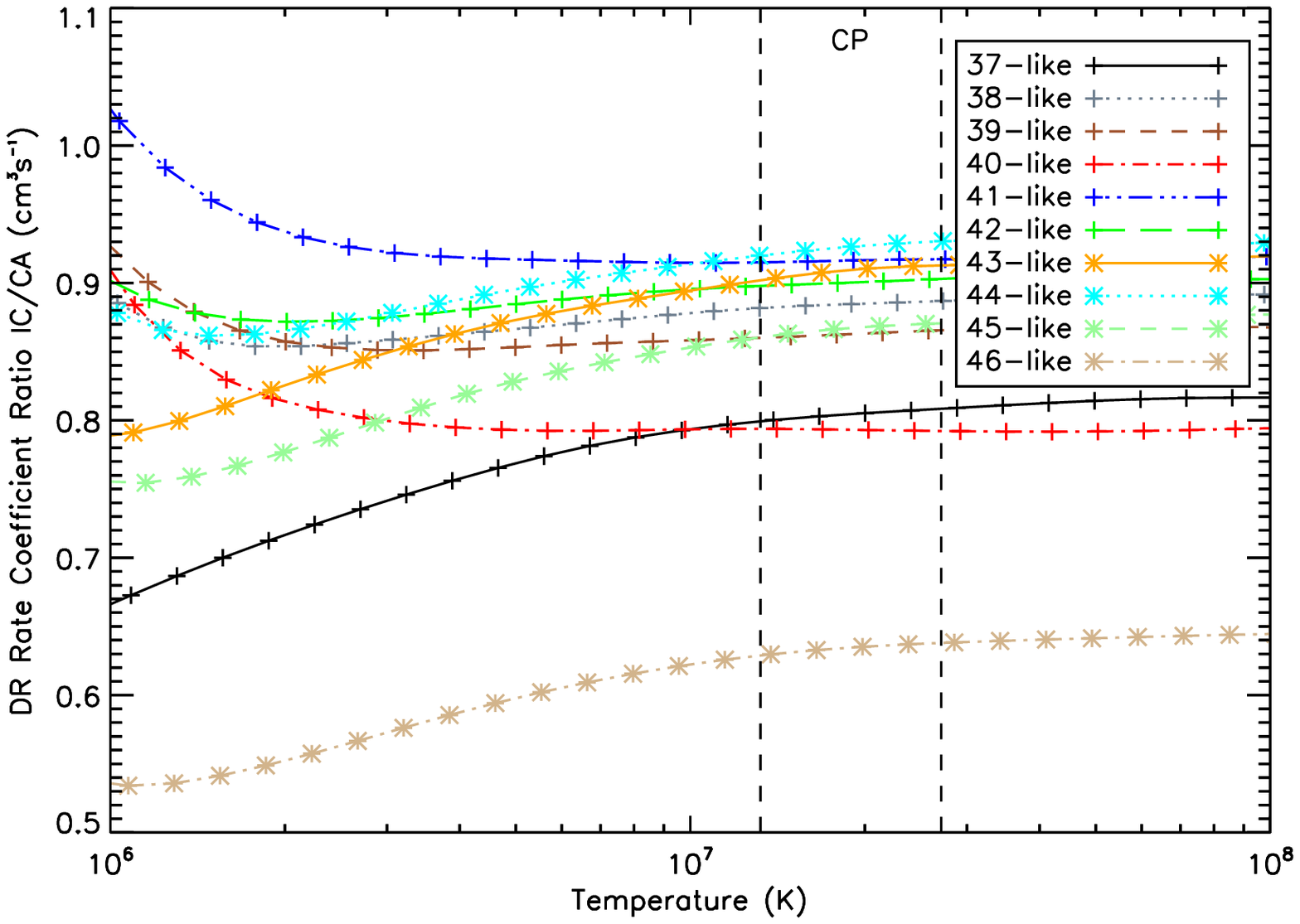}
\caption{Ratio of the total IC DR rate coefficients for 4--4 to the CA rate coefficients.}
\label{fig:ratio44}
\end{centering}
\end{figure}

4--4 is the largest contributor to the total recombination rate coefficient, constituting roughly two thirds of the total
for all of the $4d^{q}$ ions. An example of the configurations
included for the 4--4 core excitation is given for 40-like in Table \ref{table:configs40like}. In Figure \ref{fig:hicr44}
we have plotted the total 4--4 DR rate coefficient for 37- to 46-like calculated in IC. As with 3--4, we 
indicate the range of peak abundances for these ions with the CP region limited by two vertical 
dashed lines. Very little variation is seen in the DR rate coefficient in the CP region implying 
insensitivity to the atomic structure of the problem. For 37-like, 4--4 contributes $\sim{60}$\% at 
peak abundance temperature ($2.7\times{10}^{7}$K). As the $4d$ shell fills, the 4--4 contribution 
increases to 73\% at peak abundance temperature ($2.3\times{10}^{7}$K) for 40-like. By 43-like, 
4--4 contributes 76\% to the total recombination rate coefficient at peak abundance temperature
($1.8\times{10}^{7}$K). Lastly, by 46-like, the 4--4 contribution decreases slightly, contributing
$\sim{70}$\% at peak abundance temperature ($1.3\times{10}^{7}$K) to the total. The total CA DR 
rate coefficients for the 4--4 core excitation are quite similar to their IC counterparts in that
they do not change much between different $4d$ ions. Therefore, instead of plotting the CA totals, 
we opt to plot the ratio of the IC results to the CA results in Figure \ref{fig:ratio44}. It can 
be seen that the ratio deviates strongly from unity in the CP region for all ions. This deviation
becomes even larger towards low temperatures.

\subsection{4--5} \label{corex45}
\begin{figure}
\begin{centering}
\includegraphics[width=85mm]{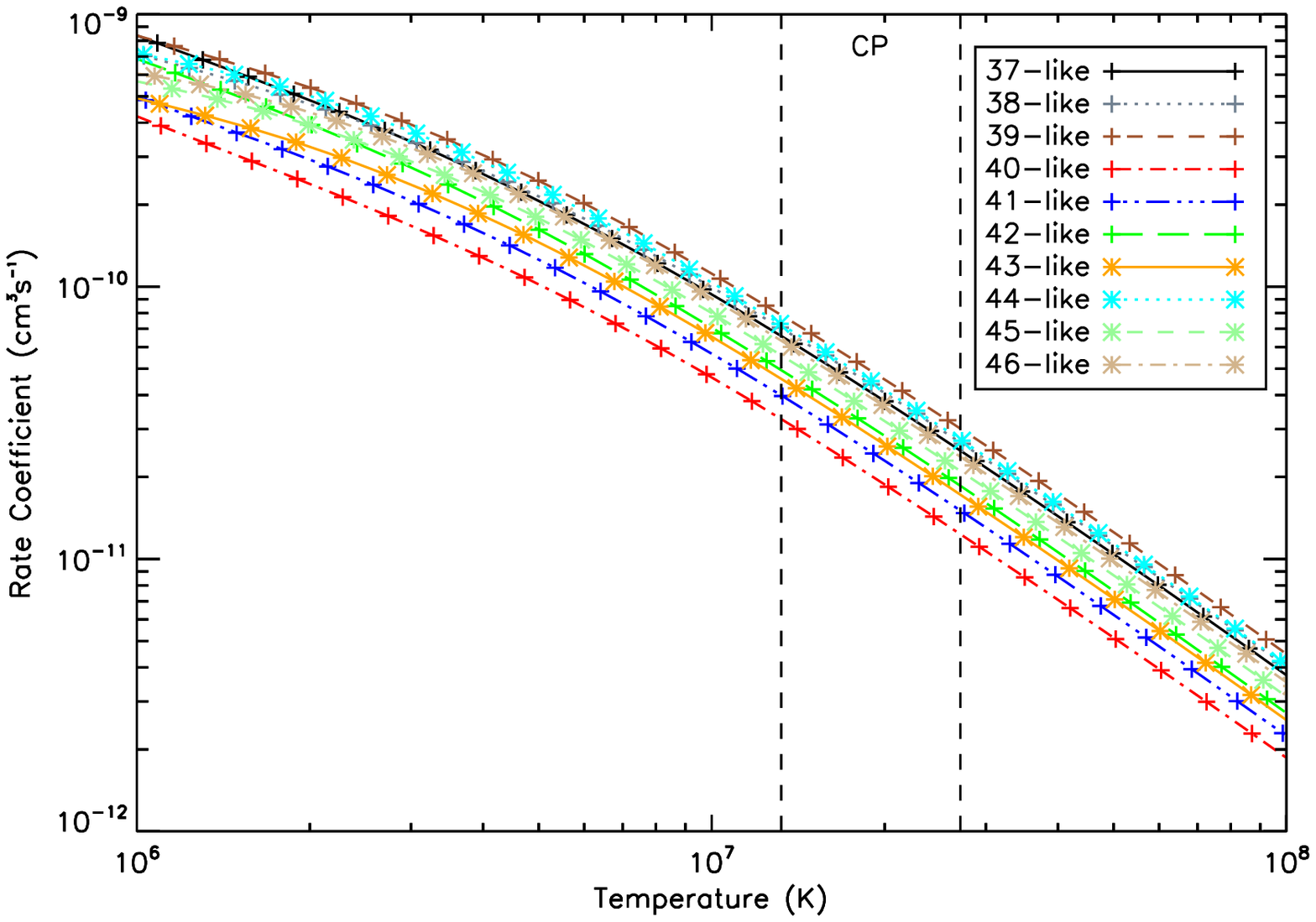}
\caption{Total DR rate coefficients for the 4--5 core excitation for 37- to 46-like, calculated
in IC.}
\label{fig:hicr45}
\end{centering}
\end{figure}

The 4--5 core excitation is the second largest contributor to the total recombination rate 
coefficient. In Figure \ref{fig:hicr45} we have plotted the total 4--5 DR rate coefficients 
calculated in IC. As with Figure \ref{fig:hcar34}, we have indicated the CP region with two 
vertical dashed lines. For 37-like, 4--5 contributes $\sim{20}$\% at peak abundance temperature. 
As we move along the isonuclear sequence, this contribution fluctuates slightly, ranging from a 
minimum of 15\% for 40-like at peak temperature, to a maximum of 30\% for 39-like. For 46-like, 
4--5 contributes 25\% at peak abundance temperature. 

As noted previously, 37-, 38-, and 46-like 4--5 were calculated including
contributions from $4p$ and $4d$, as well as one-up and one-down configurations, and are hence more
``complete'' calculations than the other ions. Furthermore, 39-, 44-, and 45-like include contributions from
$4p$ and $4d$, but do not include one-up one-down mixing. Upon close inspection of Figure \ref{fig:hicr45} it
can be seen that 40-, 41-, 42, and 43-like are separated from the other 4--5 rate coefficient 
curves. In addition, despite 37-, 38-, and 46-like including one-up one-down mixing, these 
DR rate coefficients have similar values in the CP region to those excluding these mixing 
configurations. This implies that the dominant mixing effect on the total 4--5 DR rate coefficient 
comes from the mixing of single $4p$ and $4d$ promotions, rather than the double promotions from 
one-up one-down mixing. 

\begin{figure}
\begin{centering}
\includegraphics[width=85mm]{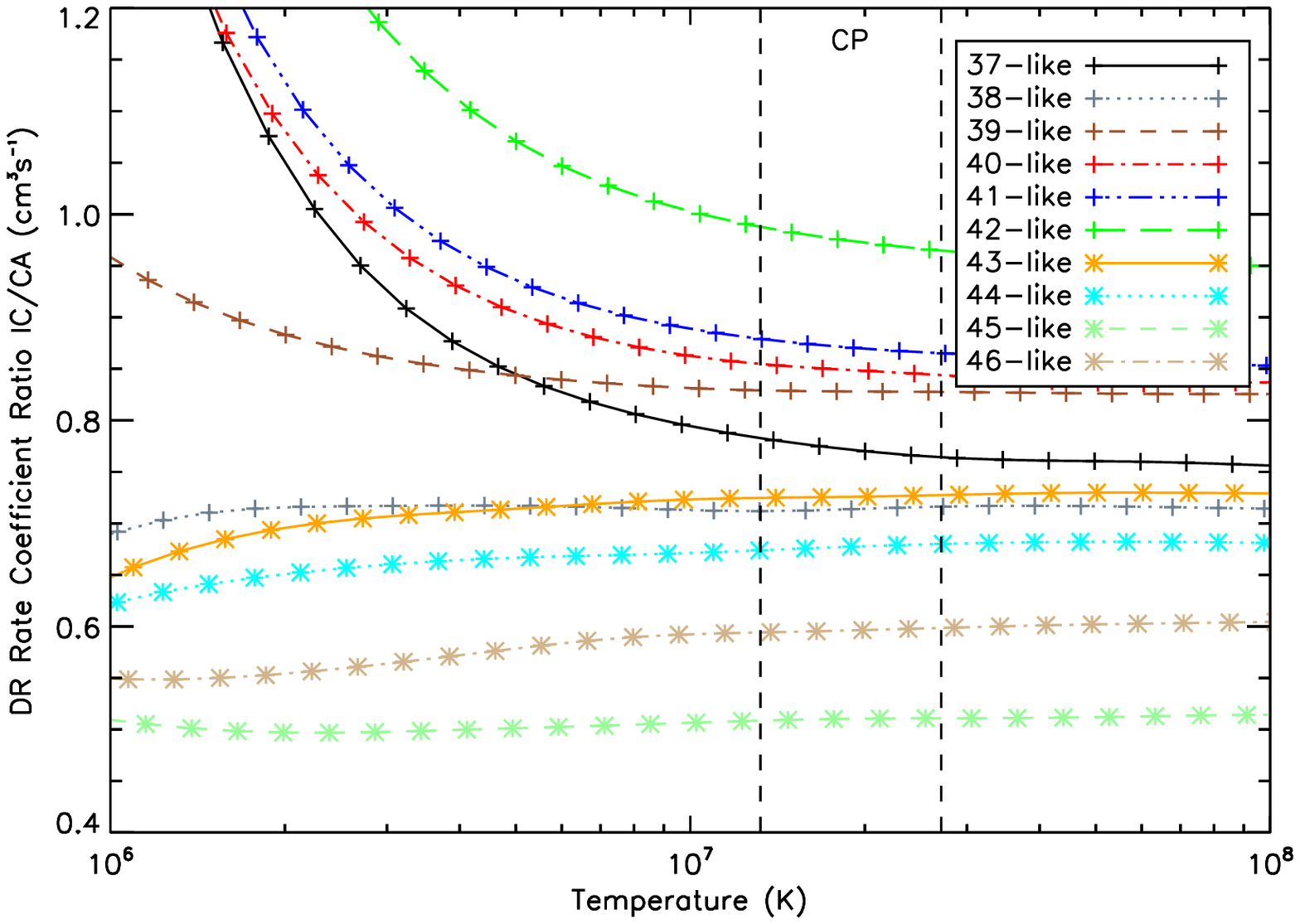}
\caption{Ratio of the total IC DR rate coefficients for 4--5 to the CA rate coefficients.}
\label{fig:ratio45}
\end{centering}
\end{figure}

As with 4--4, the CA DR rate coefficients are quite similar for all of the $4d$ ions, making comparison with
their IC counter parts difficult. In Figure \ref{fig:ratio45} we have plotted the ratio of the IC results to the 
CA results for the 4--5 core excitation. Again, large deviations are seen towards low temperatures.

\subsection{4--6}
\begin{figure}
\begin{centering}
\includegraphics[width=85mm]{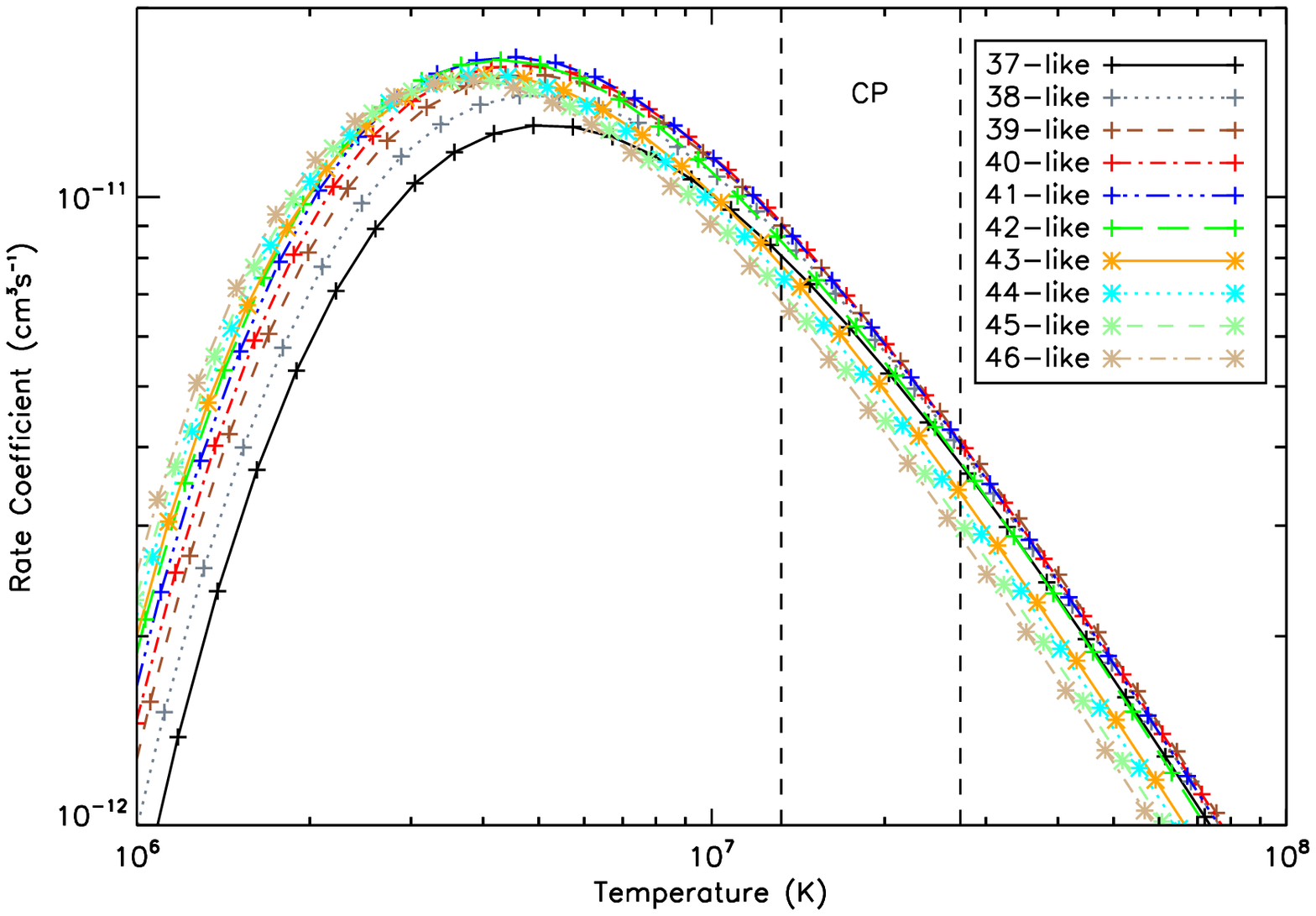}
\caption{Total DR rate coefficients for the 4--6 core excitation for 37- to 46-like, calculated
in CA.}
\label{fig:hcar46}
\end{centering}
\end{figure}

As with 3--4, the 4--6 core excitation has been calculated in CA only. 4--6 contributes very little 
to the total recombination rate coefficient, and is just included for completeness. 4--6 
constitutes just 3\% for 37- and 46-like at peak abundance temperature. We have plotted the total 
4--6 DR rate coefficients for 37- to 46-like in Figure \ref{fig:hcar46}. Little variation is seen 
for 37- to 46-like in terms of the peak temperature, which ranges from $4-5\times{10}^{6}$K. 

\section{DR threshold sensitivity}
The density and positioning of resonances near threshold can have drastic effects on the DR rate
coefficient at low temperatures. If the density of resonances is sufficiently large, then the low
temperature DR rate coefficient becomes insensitive to shifts around the threshold. To demonstrate this, 
we considered the case of 41-like for the 4--4 core excitation. We shifted the ground state of 41-like 
(ionization limit of 42-like) by 0.01, and then by 0.1 Ryd. This lowers all resonance positions. Then 
we compared the resultant total DR rate coefficients to the unshifted case. We have tabulated these DR rate 
coefficients in Table \ref{table:enhance}. It can be seen that despite these shifts, the total DR rate coefficients 
vary only slightly by 6-10\% at the lowest temperature considered. While not shown in the present paper, 
we have checked the sensitivity of the level resolved DR rate coefficient to shifts at low temperature for all $4d$ 
ions. For 38-, and 44- to 46-like, we found that under the 0.1 Ryd shift the DR rate coefficient for the 
lowest temperatures differs from the unshifted value by 33--88\%. For 37-, and 39-43-like, the difference
under the 0.1 Ryd shift is much smaller, with the lowest temperature DR rate coefficient differing
by 0--15\% from the unshifted value.

\section{Relativistic Configuration Mixing}
\begin{figure}
\begin{centering}

\includegraphics[width=85mm]{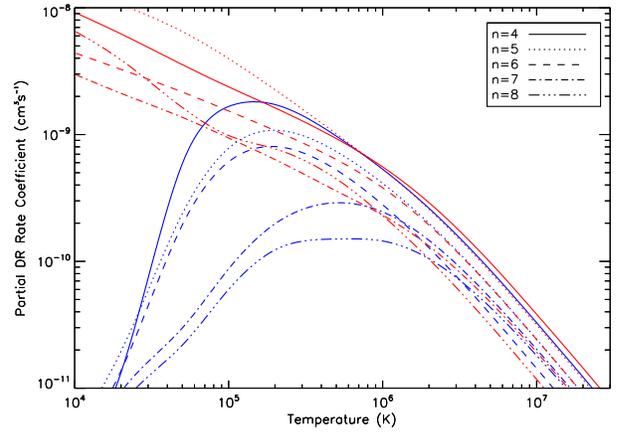}
\caption{Plot of partial DR rate coefficients for the 4--4 core excitation for 42-like for 
recombination into $n=$ 4, 5, 6, 7, and 8, calculated in IC (red curves) and CA (blue curves).}
\label{fig:mixingpartial}
\end{centering}
\end{figure}

As mentioned in the introduction Badnell~\etal \cite{badnell2011b} considered the effect of mixing 
in $4d$-shell ions of tin, presenting the case of Sn$^{10+}$ as an example. They compared
three line strength calculations for Sn$^{10+}$, calculated including promotions for $4p-4d$ 
only, $4d-4f$ only, and $4p-4d$ and $4d-4f$ promotions together. The combined case produced significantly
larger line strengths than the individual promotion cases.  Mixing is consistent across
isoelectronic sequences, meaning that mixing effects observed in tin ions will also be observed 
in tungsten ions. As seen in Sections \ref{corex44} and \ref{corex45} the ratio of the IC to CA
DR rate coefficients for the 4--4 and 4--5 core excitations differed greatly from unity, indicating
strong mixing effects. This can be more easily seen through consideration of the partial DR rate
coefficients. We consider the case of the 4--4 core excitation for 42-like, and examine recombination
into $n=$ 4, 5, 6, 7, and 8 in both IC and CA. We plot these rate coefficients in Figure 
\ref{fig:mixingpartial}. 

In all cases it can be seen that the IC and CA results diverge at low temperatures. This is simply
due to the lack of resonances at threshold in the CA calculation. We focus our discussion on temperatures
$>1\times{10}^{6}$K. The smallest differences between the IC and CA partials are observed for recombination
into $n=4$ and $n=7$. The $n=4$ IC partial is larger than the CA partial by 18\%, while the $n=7$ IC partial
is smaller than its CA counterpart by 4\%. For recombination into $n=5$ and 6, the IC partial is larger than
the CA partial by 30 and 75\% respectively. Lastly, for recombination into $n=8$, the IC partial is 
smaller than its CA counterpart by 42\%.

\section{Ionization balance}
\begin{figure*}
\begin{centering}
\includegraphics[width=170mm]{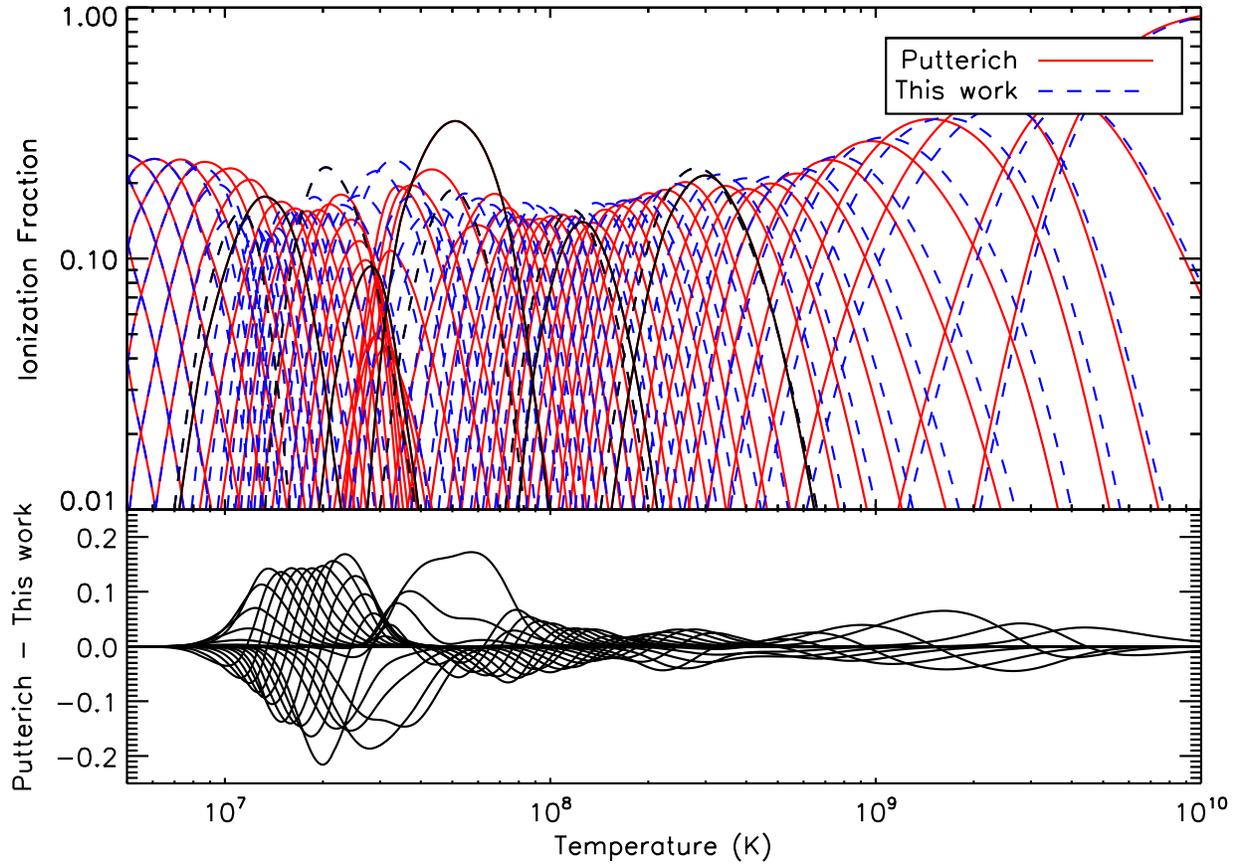}
\caption{Plot of the zero density ionization fractions for tungsten. The red-solid fraction was 
calculated using the recombination rate coefficients of P\"{u}tterich~\etal \cite{putterich2008a}, and 
the ionization rate coefficients of Loch~\etal \cite{loch2005a}. The blue-dashed fraction was 
calculated in the same way, but the recombination rate coefficient data for 00- to 46-like was 
replaced with our data from the present work, and Preval~\etal \cite{preval2016a,preval2017a}. 
From right to left, the black parabolas indicate the 10-, 18-, 28-, 36-, and 46-like ionization 
fractions.}
\label{fig:ionfrac}
\end{centering}
\end{figure*}

We now consider the impact of our calculations on the coronal-approximation ionization fraction of tungsten. 
We consider this approximation for simplicity. This is done by first calculating the ionization fraction using the 
recombination rate coefficients of P\"{u}tterich~\etal \cite{putterich2008a} and the ionization rate coefficients 
of Loch~\etal \cite{loch2005a}. We then compare this with a similar calculation, where we replace the DR+RR 
rate coefficients with the values reported here. We plot the two ionization fractions in Figure \ref{fig:ionfrac}, 
along with the arithmetic difference between the P\"{u}tterich~\etal and current ionization fractions. Note that 
for our discussion of these ionization fractions, we have omitted the J\"{u}ttner correction described in 
Section \ref{theorysec}.

The most striking result emerging from this plot is the close grouping of the peak abundance temperatures for 
the $4d$ ions. In the present ionization fractional abundance, the envelope of peak temperatures is 0.97-1.6keV 
for 46- to 37-like, whereas in the P\"{u}tterich~\etal ionization fraction, the same stages peak at higher 
temperatures and cover a wider range, 1.1-­2.3keV. The new rates result in a significant narrowing of the fractional 
abundances, almost by a factor of 2 at the FWHM value. The stages adjacent to this group, 47-like to 50-like, also 
show a narrowing effect (at a smaller 20\% level) which shows the influence of neighbouring ionization stages.

In addition, the peak fraction using our data ranges from 0.13 to 0.17, whereas the P\"{u}tterich~\etal et al peak 
fraction ranges from 0.1 to 0.2. The consistency of the peak fraction and the narrow temperature range in which 
the $4d$-shell ions are most abundant can potentially offer an accurate plasma diagnostic. Observation of spectral 
emission originating from a $4d$ ion would tightly constrain the temperature of the originating plasma. The 
temperature region of these ionization stages will be of particular interest to the fusion community as the pedestal 
temperature at ITER is expected to be in the 1-3keV range.

In Table \ref{table:peaktemp} we give revised peak abundance temperatures and fractions when using our rate 
coefficients compared to using P\"{u}tterich~\etal's data, spanning 01-like to 46-like. The effect on the ionization 
balance of replacing the ten $4d$-shell DR rates, 46-like to 37-like are clearly seen between $1-2\times{10}^{7}$K. 
The extent of the effect of the new rates extends beyond those replaced and is present at the $10^{-3}$ level up 
to 28-like (W$^{46+}$) and down to 53-like (W$^{21+}$). This accounts for the numerical difference between Table 
\ref{table:peaktemp} in the present work, and Table 7 of Preval~\etal \cite{preval2017b} for some of the common 
ionization stages. Due to the effect on the abundances of adjacent stages because of replacing these rates some 
caution should be exercised when using the abundances below W$^{28+}$.

\section{Comparison with P\"{u}tterich~\etal}
\begin{figure}
\begin{centering}
\includegraphics[width=85mm]{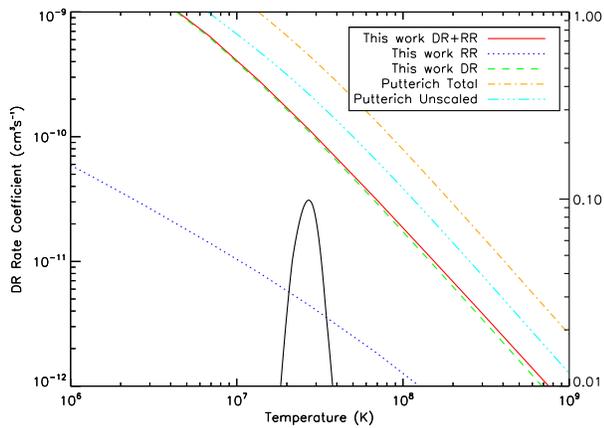}
\caption{Comparison of the present recombination rate coefficients, split into DR (green-dashed),
RR (blue-dotted), and the total of these (solid-red) with P\"{u}tterich~\etal$\!\!$'s scaled (orange-dot) 
and unscaled (cyan dash triple dot) results for 37-like. The solid black parabola indicates the
peak abundance fraction for 37-like.}
\label{fig:puttprev37}
\end{centering}
\end{figure}

\begin{figure}
\begin{centering}
\includegraphics[width=85mm]{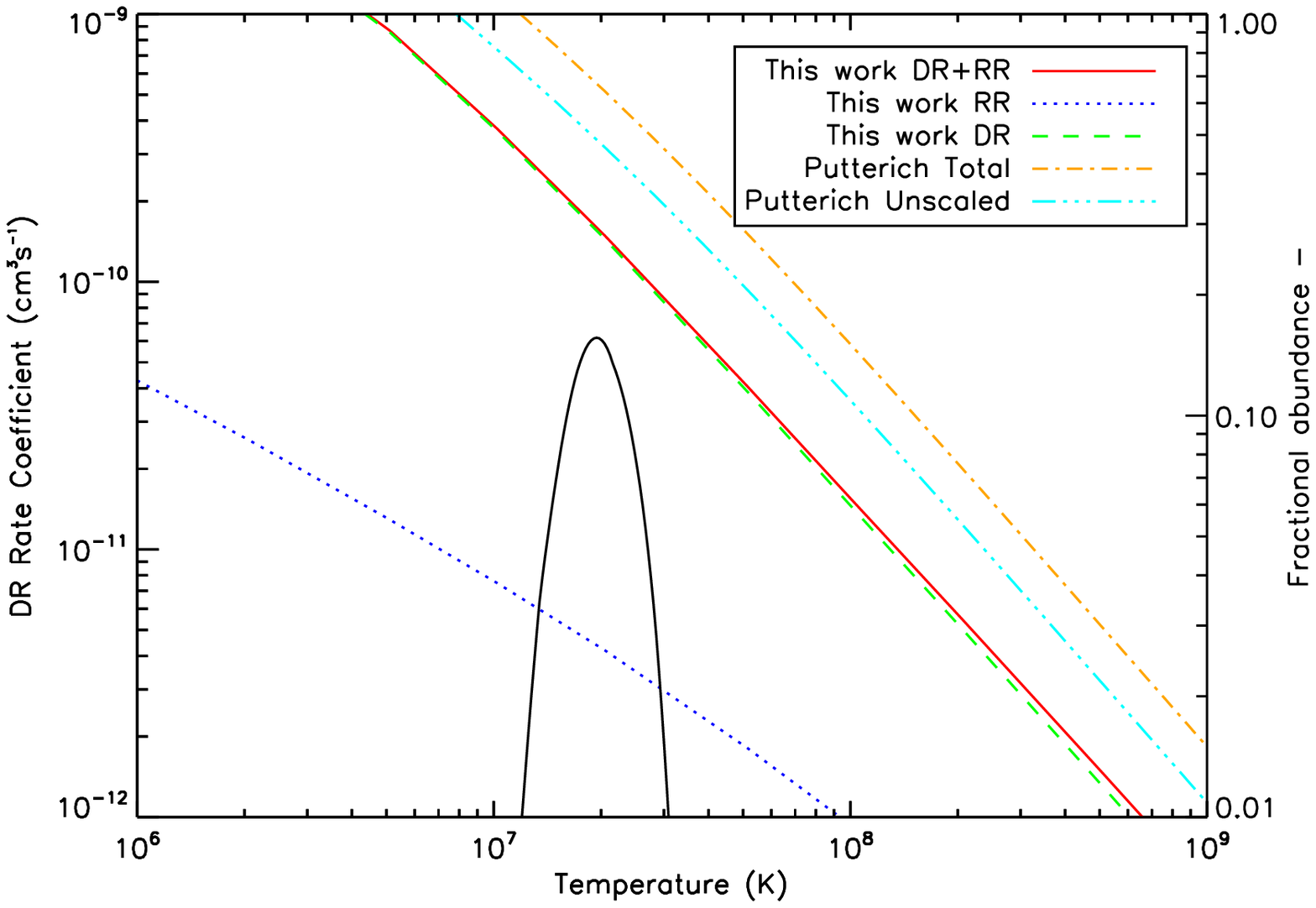}
\caption{Same as Figure \ref{fig:puttprev37}, but for 42-like.}
\label{fig:puttprev42}
\end{centering}
\end{figure}

\begin{figure}
\begin{centering}
\includegraphics[width=85mm]{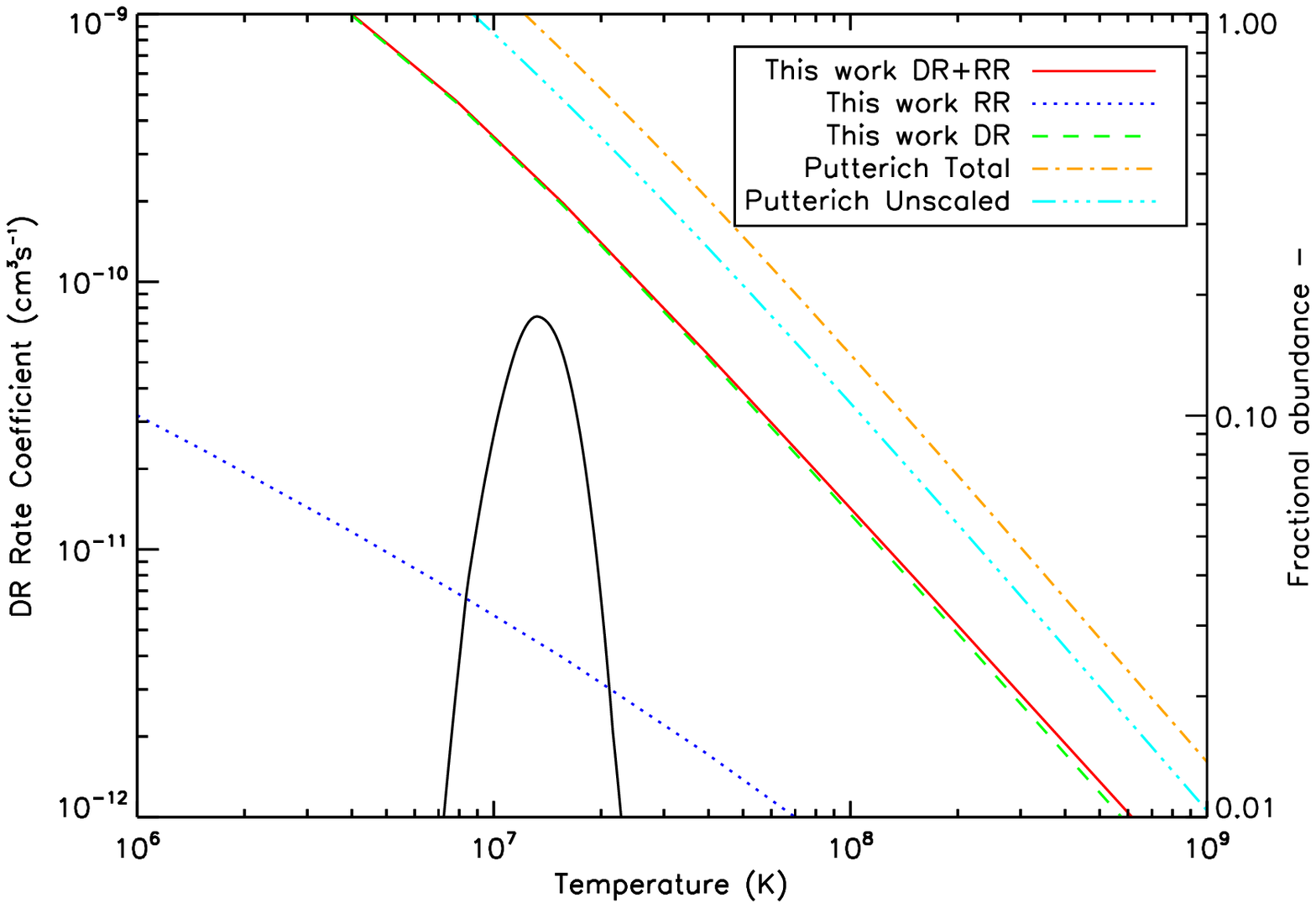}
\caption{Same as Figure \ref{fig:puttprev37}, but for 46-like.}
\label{fig:puttprev46}
\end{centering}
\end{figure}

The agreement between our total recombination rate coefficients and P\"{u}tterich~\etal$\!\!$'s 
\cite{putterich2008a} is very poor for all $4d$ ions, with their rates being consistently higher
than the present work. In Figures \ref{fig:puttprev37}, \ref{fig:puttprev42}, and 
\ref{fig:puttprev46} we have plotted our total recombination rate coefficients for 37-like, 
42-like, and 46-like respectively, along with P\"{u}tterich~\etal$\!$'s scaled and unscaled results. 
The differences between the present work and P\"{u}tterich~\etal$\!\!$'s are fairly constant. 
The scaled results are a factor 3-4 larger than the present work at peak abundance
temperature, while the unscaled results are a factor 2-2.5 larger.

\section{Comparison with other works}
Unlike the ionization states considered in Preval~\etal \cite{preval2017a}, very few detailed 
calculations exist for the $4d$-shell ions of tungsten. To date, detailed calculations have been
performed for 37-, 39-, 45-, and 46-like. We compare each of these calculations with the present
results.

\subsection{46-like}
\begin{figure}
\begin{centering}
\includegraphics[width=85mm]{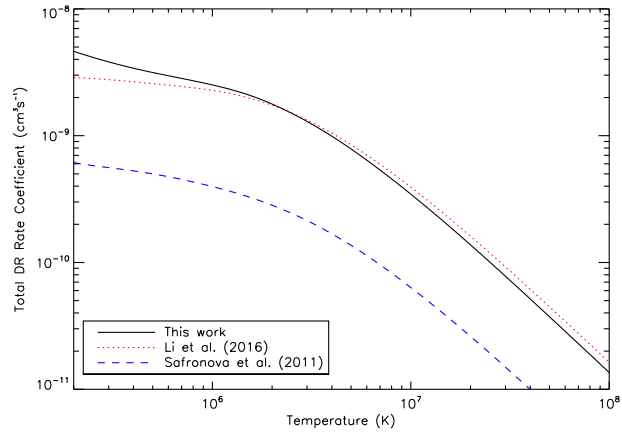}
\caption{Total DR rate coefficients for 46-like as calculated in the present work (black solid
line, by Li~\etal \cite{li2016a} (red dotted line), and by Safronova~\etal \cite{usafronova2011a} (blue dashed line).}
\label{fig:46rates}
\end{centering}
\end{figure}

Like other closed shell ions, 46-like is an important ion for plasma diagnostic purposes. As well
as the present work, DR rate coefficients for this ion have been calculated by Safronova~\etal 
\cite{usafronova2011a} using {\sc hullac} and Li~\etal \cite{li2016a} using {\sc fac}. We have plotted the present total DR rate
coefficients, along with those calculated by Safronova~\etal and Li~\etal in Figure 
\ref{fig:46rates}. Our and Li~\etal$\!\!$'s DR rate coefficients are in relatively good agreement, 
differing by $\sim{0-10}\%$ between $\sim{9}\times{10}^{5}-6\times{10}^{6}$K. This difference 
gradually increases to 20\% by $1\times{10}^{8}$K. 

\begin{figure}
\begin{centering}
\includegraphics[width=85mm]{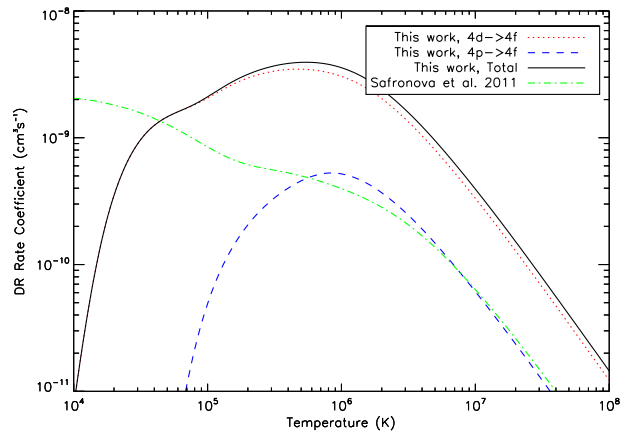}
\caption{Comparison of total CA DR rate coefficients for the 46-like 4--4 core excitation to
the total DR rate coefficients of Safronova~\etal \cite{usafronova2011a}. We also include the constituent 4d-4f and
4p-4f contributions to the 4--4 total.}
\label{fig:46rates2}
\end{centering}
\end{figure}

Our total DR rate coefficients are consistently larger than Safronova~\etal$\!\!$'s by a factor $\sim{8}$ 
over all temperatures. We note that the DR rate coefficients calculated by Safronova~\etal neglects
$4p$ excitations in the $N$- and $(N+1)$-electron targets. However, the contribution from $4p$ alone
is not enough to account for the large difference between our and Safronova~\etal$\!\!$'s rate
coefficients. We demonstrate this in Figure \ref{fig:46rates2}, where we have plotted the total DR-rate 
coefficients calculated in CA and IC for the 4--4 core excitation, along with their respective 
$4p$ and $4d$ contributions. At maximum, $4p$ contributes 15\% to the 4--4 CA total at high 
temperatures, while the rest is provided by $4d$. Because our results are in relatively good 
agreement with those calculated by Li~\etal, the large difference between our DR rate coefficients 
and Safronova~\etal$\!\!$'s appears to be due to some systematic problem/omission in the authors' calculation. 

We note that large differences between our results and Safronova~\etal$\!\!$'s are not limited to the
present work. For example, Preval~\etal \cite{preval2016a} noted a difference of $\sim{47}$\% between
their work and that of Safronova~\etal \cite{usafronova2009b} for 10-like tungsten. However, 
comparison of Preval~\etal$\!\!$'s 10-like results with Behar~\etal \cite{behar1999a} (calculated with 
{\sc hullac})showed differences of $<10$\%. More recently, a calculation by Li~\etal \cite{li2016a} 
for 10-like showed differences of 30-40\% with Safronova~\etal \cite{usafronova2009b} over
a wide temperature range while showing good agreement with Preval~\etal at peak abundance 
temperature. In addition, Preval~\etal \cite{preval2017b} recently compared their calculated 28- and
29-like DR rate coefficients to those presented by Safronova~\etal 
\cite{usafronova2012b,usafronova2015a}. As in the previous case, large systematic differences were
observed of 27\% and a factor 11 for 28- and 29-like respectively. Excellent agreement was
seen between Preval~\etal and Behar~\etal \cite{behar1999b} for 28-like with differences of 
$\sim{7}$\% at peak abundance temperature. For 29-like, good agreement between Preval~\etal and 
Kwon~\etal \cite{kwon2016a} was seen with differences of 23\% at peak abundance temperature.
Without repeating Safronova~\etal$\!\!$'s calculations (which is beyond the scope of this work), the
origin of the aformentioned systematic difference is unknown.

\subsection{45-like}
\begin{figure}
\begin{centering}
\includegraphics[width=85mm]{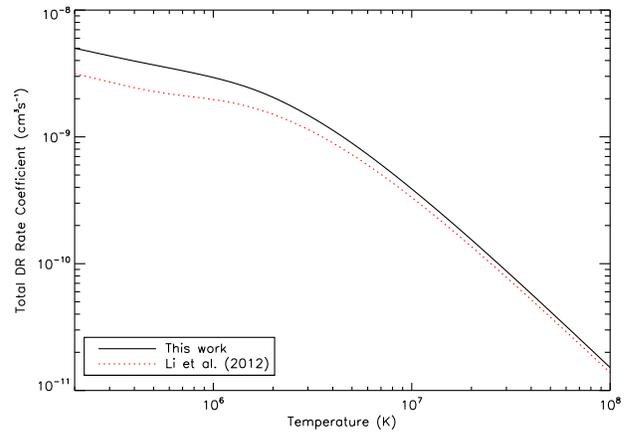}
\caption{Total DR rate coefficients for 45-like as calculated in the present work (black solid
line, and by Li~\etal \cite{li2012a} (red dotted line).}
\label{fig:45rates}
\end{centering}
\end{figure}

Only one calculation has been done for 45-like tungsten. Li~\etal \cite{li2012a}
used {\sc fac} to calculate DR rate coefficients for this ion. In Figure \ref{fig:45rates} we have 
plotted our DR rate coefficeints for 45-like, along with Li~\etal$\!\!$'s results. Good agreement is 
seen over the entire temperature range, with the best agreement seen at high temperatures. At 
$10^{8}$K our DR rate coefficients are smaller than Li~\etal$\!\!$'s results, differing by ~9\%. As we move towards peak 
abundance temperature ($2.9\times{10}^{7}$K), this difference increases slightly to 11\%. The 
maximum difference between our and Li~\etal$\!\!$'s DR rate coefficients is ~40\%, occuring at 
$\sim{10}^{5}$K. This difference is unlikely to be due to the positioning of resonances at 
low temperature, as we have previously shown that the DR rate coefficient is insensitive to
positioning at threshold. 

\subsection{39-like}
\begin{figure}
\begin{centering}
\includegraphics[width=85mm]{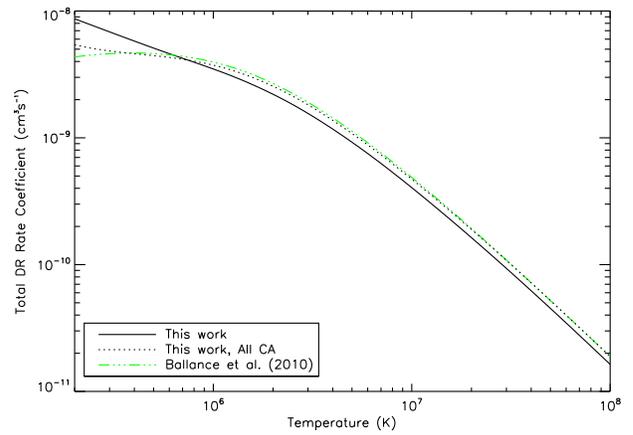}
\caption{Total DR rate coefficients for 39-like as calculated in the present work (black solid
line, and by Ballance~\etal \cite{ballance2010a}. (green triple dot-dashed line). We also include our CA result for 
this ion (black dotted).}
\label{fig:39rates}
\end{centering}
\end{figure}

The only calculation available for 39-like was done by Ballance~\etal \cite{ballance2010a}, who 
used a CA distorted wave (CADW) code, named {\sc dracula} (Griffin~\etal \cite{griffin1985a}) to
calculate DR for $\Delta{n}=0,1,2$ processes. In Figure \ref{fig:39rates} we have plotted 
Ballance~\etal$\!\!$'s calculation, along with our total DR rate coefficient calculated in both IC and 
CA. The shape of all three curves matches very well from high temperatures down to $\sim{10^{6}}$K.
The present CA results are in near perfect agreement with Ballance~\etal$\!\!$'s, differing by $<0.1$\% 
for temperatures $>{10^{7}}$K. As Ballance~\etal$\!\!$'s 
calculation mirrors ours at high temperature, this shows our result is reliable. The difference 
between the present IC result and the CADW result is fairly constant between $\sim{10^{7}-10^{8}}$K, 
differing by 15-16\%. Differences between all three curves appear for temperatures $<{10^{6}}$K. 
It has been shown previously that the low temperature DR rate coefficient in IC is insensitive to 
resonance positioning due to the desity of resonances near threshold. However, in the case of CA,
there are far fewer resonances than in IC, meaning that any differences will be dependent upon
the quality of the structure calculation.

\subsection{37-like}
\begin{figure}
\begin{centering}
\includegraphics[width=85mm]{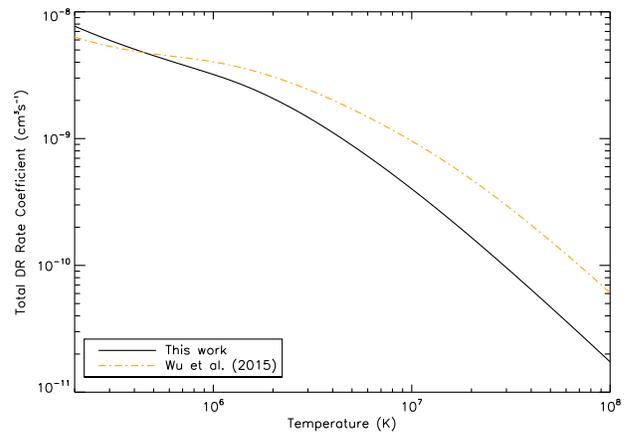}
\caption{Total DR rate coefficients for 37-like as calculated in the present work (black solid
line, and by Wu~\etal \cite{wu2015b} (orange dot-dashed line).}
\label{fig:37rates}
\end{centering}
\end{figure}

37-like is the simplest $4d$-shell ion with only a single $4d$ electron. Interestingly, this ion has
been considered twice by Wu~\etal \cite{wu2015a,wu2015b}. For the purpose of this comparison, we 
compare with the results of the latest analysis by the authors, where they calculated DR rate 
coefficients for 37-like using {\sc fac}. In Figure \ref{fig:37rates} we have plotted our total DR rate 
coefficients for 37-like, along with Wu~\etal$\!\!$'s \cite{wu2015b} results. Large differences can be seen between our 
results and Wu~\etal$\!\!$'s for temperatures $>5\times{10}^{5}$K, with the differences gradually 
getting larger with increasing temperature. At peak abundance temperature, Wu~\etal$\!\!$'s DR rate 
coefficients are larger than ours by a factor $\sim3$. A similar discrepancy was described in
Preval~\etal \cite{preval2017a}, where other ions considered by Wu~\etal were also much larger than
our own results. It was noted that the discrepancy between the two results could be reduced by
removing the core-rearrangement configurations from Preval~\etal$\!\!$'s inner shell DR calculations, 
hence, reducing high temperature Auger suppression. However, as noted earlier, the inner shell DR
core excitation 3--4 only contributes 12\% at peak abundance, and ~14\% at higher temperatures, 
meaning this cannot be the source of the discrepancy. We also note in our calculations that the high
temperature DR rate coefficient does not vary much from ion to ion, and we are in close agreement with
Ballance~\etal \cite{ballance2010a} for 39-like.

\section{Conclusions}
We have presented a series of DR and RR rate coefficient calculations for the $4d$-shell ions of
tungsten spanning 37- to 46-like. The present work is the first consideration of the $4d^{q}$
($q=1-10$) ions of tungsten. The rich level structure
of the $4d^{q}$ ions, particularly for half-open ions, renders the level-resolved
DR rate coefficient insensitive to shifts in the ground state energy due to a high density of resonances
close to threshold. 

Very few detailed calculations were available to compare the present results with. 
Relatively good agreement was seen between the present work, and the calculations performed by 
Ballance~\etal \cite{ballance2010a} for 39-like, and Li~\etal \cite{li2012a} for 45- and 46-like at peak 
abundance temperature. Very poor agreement was seen between the present work, and the 
calculations performed by Wu~\etal \cite{wu2015a} and Safronova~\etal \cite{usafronova2011a}. 
Unfortunately, it has still not been possible to pin down the reason for the difference between our 
calculations and Safronova~\etal$\!\!$'s. We found that the present results were consistently smaller than the scaled 
ADPAK data of P\"{u}tterich~\etal \cite{putterich2008a} by a factor 3-4, and a factor 2-2.5 when compared to the 
unscaled data. 

In terms of practical applications, revision of the $4d$-shell atomic data has shifted
the peak abundance temperatures of these ions to values comparable to the projected pedestal 
temperature of ITER. This is significant, as it implies the lower ionization state tungsten ions (most
notably the $4f$-shell ions) may not be as important as the higher ionization states. 

With this paper, only 27 ionization stages remain to be calculated for \textit{The Tungsten Project}.
These stages span the $4f$-shell, and then continue into the $n=5$ and $n=6$ shells. Given the 
computational difficulty of the $4f$ shell, we opt to next explore the lanthanide series of ions from 
61-like to singly-ionized tungsten. We will then complete \textit{The Tungsten Project} with a final 
paper covering the $4f$ ions of tungsten, from 47- to 60-like.

\ack
SPP, NRB, and MGOM acknowledge the support of EPSRC grant EP/1021803
to the University of Strathclyde. All data calculated as part of this work are 
publicly available on the OPEN-ADAS website https://open.adas.ac.uk.

\section*{References}
\bibliography{simonpreval}
\newpage

\clearpage


\begin{table}
\caption{Example configuration set for the 40-like 4--4 core excitation. Configurations marked with
an * are mixing configurations as determined by the one-up one down rule.}
\begin{tabular}{@{}llcccccc}
\hline
\hline
N-electron              & $(N+1)$-electron \\
\hline
$4p^{6} 4d^{4}         $ & $4p^{6} 4d^{5}         $ \\
$4p^{6} 4d^{3} 4f      $ & $4p^{6} 4d^{4} 4f      $ \\
$4p^{5} 4d^{5}         $ & $4p^{6} 4d^{3} 4f^{2}  $ \\
$4p^{5} 4d^{4} 4f      $ & $4p^{5} 4d^{6}         $ \\
$4p^{6} 4d^{2} 4f^{2}* $ & $4p^{5} 4d^{5} 4f      $ \\
$4p^{4} 4d^{6}*        $ & $4p^{5} 4d^{4} 4f^{2}  $ \\
                         & $4p^{4} 4d^{7}*        $ \\
                         & $4p^{4} 4d^{6} 4f*     $ \\
                         & $4p^{6} 4d^{2} 4f^{3}* $ \\
\hline
\hline
\label{table:configs40like}
\end{tabular}
\end{table}

\begin{table*}
\caption{Comparison of peak abundance temperatures and fractions as calculated using P\"{u}tterich~\etal$\!\!$'s data \cite{putterich2008a}, and
P\"{u}tterich~\etal$\!\!$'s data with 01- to 36-like replaced with our data. The ionization rate coefficients originate from Loch~\etal \cite{loch2005a}. Note $[x]=10^{x}$.}
\begin{tabular}{@{}lccccccc}
\hline
\hline
Ion-like & Charge & Putt $T_{\mathrm{peak}}$ & Putt $f_{\mathrm{peak}}$ & This work $T_{\mathrm{peak}}$ & This work $f_{\mathrm{peak}}$ & $\Delta{T}$\% & $\Delta{f}$\% \\
\hline
01-like & W$^{73+}$ & 3.88[+9] & 0.440 & 4.06[+9] & 0.426 &  4.72 & -3.04 \\
02-like & W$^{72+}$ & 2.29[+9] & 0.442 & 2.46[+9] & 0.405 &  7.66 & -8.39 \\
03-like & W$^{71+}$ & 1.48[+9] & 0.360 & 1.64[+9] & 0.363 &  11.2 &  0.94 \\
04-like & W$^{70+}$ & 9.66[+8] & 0.294 & 1.06[+9] & 0.304 &  9.70 &  3.41 \\
05-like & W$^{69+}$ & 7.17[+8] & 0.247 & 7.67[+8] & 0.255 &  6.97 &  3.30 \\
06-like & W$^{68+}$ & 5.72[+8] & 0.218 & 5.97[+8] & 0.228 &  4.37 &  4.35 \\
07-like & W$^{67+}$ & 4.76[+8] & 0.199 & 4.84[+8] & 0.213 &  1.64 &  6.70 \\
08-like & W$^{66+}$ & 4.03[+8] & 0.189 & 3.98[+8] & 0.209 & -1.34 &  10.3 \\
09-like & W$^{65+}$ & 3.46[+8] & 0.195 & 3.32[+8] & 0.216 & -3.79 &  11.1 \\
10-like & W$^{64+}$ & 2.99[+8] & 0.214 & 2.82[+8] & 0.229 & -5.88 &  7.03 \\
11-like & W$^{63+}$ & 2.60[+8] & 0.202 & 2.43[+8] & 0.198 & -6.58 & -1.75 \\
12-like & W$^{62+}$ & 2.25[+8] & 0.188 & 2.11[+8] & 0.179 & -6.09 & -4.99 \\
13-like & W$^{61+}$ & 2.01[+8] & 0.182 & 1.89[+8] & 0.174 & -5.83 & -4.33 \\
14-like & W$^{60+}$ & 1.80[+8] & 0.172 & 1.70[+8] & 0.171 & -5.89 & -1.00 \\
15-like & W$^{59+}$ & 1.62[+8] & 0.163 & 1.52[+8] & 0.166 & -6.10 &  1.36 \\
16-like & W$^{58+}$ & 1.47[+8] & 0.157 & 1.38[+8] & 0.167 & -6.31 &  6.41 \\
17-like & W$^{57+}$ & 1.36[+8] & 0.139 & 1.27[+8] & 0.154 & -6.56 &  10.5 \\
18-like & W$^{56+}$ & 1.26[+8] & 0.140 & 1.16[+8] & 0.157 & -7.40 &  12.1 \\
19-like & W$^{55+}$ & 1.17[+8] & 0.147 & 1.07[+8] & 0.150 & -8.22 &  1.76 \\
20-like & W$^{54+}$ & 1.08[+8] & 0.150 & 9.83[+7] & 0.160 & -9.28 &  6.65 \\
21-like & W$^{53+}$ & 1.01[+8] & 0.148 & 9.01[+7] & 0.162 & -10.4 &  9.89 \\
22-like & W$^{52+}$ & 9.29[+7] & 0.144 & 8.33[+7] & 0.167 & -10.4 &  15.7 \\
23-like & W$^{51+}$ & 8.60[+7] & 0.143 & 7.73[+7] & 0.153 & -10.2 &  6.79 \\
24-like & W$^{50+}$ & 8.01[+7] & 0.146 & 7.14[+7] & 0.156 & -10.9 &  6.37 \\
25-like & W$^{49+}$ & 7.41[+7] & 0.159 & 6.57[+7] & 0.164 & -11.3 &  3.05 \\
26-like & W$^{48+}$ & 6.74[+7] & 0.181 & 6.02[+7] & 0.174 & -10.8 & -4.09 \\
27-like & W$^{47+}$ & 5.97[+7] & 0.136 & 5.47[+7] & 0.182 & -8.38 &  33.6 \\
28-like & W$^{46+}$ & 5.10[+7] & 0.353 & 5.03[+7] & 0.188 & -1.43 & -46.8 \\
29-like & W$^{45+}$ & 4.32[+7] & 0.227 & 4.62[+7] & 0.157 &  7.16 & -30.5 \\
30-like & W$^{44+}$ & 3.74[+7] & 0.196 & 4.21[+7] & 0.150 &  12.6 & -23.5 \\
31-like & W$^{43+}$ & 3.40[+7] & 0.194 & 3.79[+7] & 0.177 &  11.3 & -8.90 \\
32-like & W$^{42+}$ & 3.22[+7] & 0.108 & 3.30[+7] & 0.248 &  2.45 &   130 \\
33-like & W$^{41+}$ & 3.11[+7] & 0.044 & 2.94[+7] & 0.224 & -5.47 &   410 \\
34-like & W$^{40+}$ & 3.01[+7] & 0.052 & 2.59[+7] & 0.178 & -13.9 &   245 \\
35-like & W$^{39+}$ & 2.91[+7] & 0.049 & 2.29[+7] & 0.164 & -21.5 &   233 \\
36-like & W$^{38+}$ & 2.81[+7] & 0.093 & 2.05[+7] & 0.231 & -27.2 &   148 \\
37-like & W$^{37+}$ & 2.71[+7] & 0.099 & 1.88[+7] & 0.176 & -30.4 &  78.3 \\
38-like & W$^{36+}$ & 2.59[+7] & 0.118 & 1.75[+7] & 0.157 & -32.6 &  33.9 \\
39-like & W$^{35+}$ & 2.45[+7] & 0.143 & 1.63[+7] & 0.155 & -33.4 &  8.74 \\
40-like & W$^{34+}$ & 2.28[+7] & 0.180 & 1.53[+7] & 0.160 & -32.6 & -11.2 \\
41-like & W$^{33+}$ & 2.10[+7] & 0.165 & 1.45[+7] & 0.132 & -31.0 & -19.9 \\
42-like & W$^{32+}$ & 1.94[+7] & 0.156 & 1.37[+7] & 0.128 & -29.4 & -18.2 \\
43-like & W$^{31+}$ & 1.79[+7] & 0.155 & 1.31[+7] & 0.130 & -26.7 & -15.9 \\
44-like & W$^{30+}$ & 1.64[+7] & 0.159 & 1.26[+7] & 0.131 & -23.5 & -17.6 \\
45-like & W$^{29+}$ & 1.49[+7] & 0.169 & 1.20[+7] & 0.147 & -19.9 & -12.9 \\
46-like & W$^{28+}$ & 1.32[+7] & 0.177 & 1.12[+7] & 0.154 & -14.9 & -13.0 \\
\hline
\hline
\label{table:peaktemp}
\end{tabular}
\end{table*}

\begin{table*}
\caption{Comparison of total DR rate coefficients for 41-like 4--4 calculated in IC with ground 
state shifts of 0, 0.01, and 0.10 Ryd. Note $[x]=10^{x}$.}
\begin{tabular}{@{}lccccccc}
\hline
\hline
Temp (K) & No Shift & Shift 0.01 & Shift 0.10 \\
\hline
1.09[4] &  4.15[-8]  &  3.78[-8]  &  4.41[-8] \\
2.18[4] &  2.77[-8]  &  2.63[-8]  &  2.74[-8] \\
5.45[4] &  1.67[-8]  &  1.63[-8]  &  1.65[-8] \\
1.09[5] &  1.20[-8]  &  1.19[-8]  &  1.20[-8] \\
2.18[5] &  7.93[-9]  &  7.88[-9]  &  7.87[-9] \\
5.44[5] &  4.29[-9]  &  4.28[-9]  &  4.26[-9] \\
1.09[6] &  2.81[-9]  &  2.80[-9]  &  2.80[-9] \\
2.18[6] &  1.66[-9]  &  1.66[-9]  &  1.66[-9] \\
5.44[6] &  6.28[-10] &  6.27[-10] &  6.25[-10] \\
1.09[7] &  2.57[-10] &  2.57[-10] &  2.56[-10] \\
2.18[7] &  9.81[-11] &  9.80[-11] &  9.75[-11] \\
5.44[7] &  2.57[-11] &  2.57[-11] &  2.56[-11] \\
1.09[8] &  9.09[-12] &  9.09[-12] &  9.05[-12] \\
2.18[8] &  3.13[-12] &  3.13[-12] &  3.12[-12] \\
5.44[8] &  7.23[-13] &  7.23[-13] &  7.19[-13] \\
1.09[9] &  2.20[-13] &  2.20[-13] &  2.19[-13] \\
2.18[9] &  5.98[-14] &  5.97[-14] &  5.94[-14] \\
\hline
\hline
\label{table:enhance}
\end{tabular}
\end{table*}

\end{document}